\documentclass[aps,amsmath,amssymb,prd,superscriptaddress,twocolumn]{revtex4-1}

\usepackage{graphicx}
\usepackage{dcolumn}
\usepackage{bm}
\usepackage{color}
\usepackage[colorlinks=true,pdfstartview=FitV,linkcolor=blue,citecolor=blue,urlcolor=blue]{hyperref}

\everymath{\displaystyle}
\begin{document}

\newcommand{\up}[1]{$^{#1}$}
\newcommand{\down}[1]{$_{#1}$}
\newcommand{\powero}[1]{\mbox{10$^{#1}$}}
\newcommand{\powert}[2]{\mbox{#2$\times$10$^{#1}$}}

\newcommand{\dedx}{$\rm{dE/dx}$}
\newcommand{\gev}{\mbox{GeV\,}$c^{-2}$}
\newcommand{\swn}{\mbox{$\sigma_{\chi \mbox{\sc n}}$}}
\newcommand{\evr}{\mbox{eV$_{\rm nr}$}}
\newcommand{\eve}{\mbox{eV$_{\rm ee}$}}
\newcommand{\dru}{\mbox{keV$_{\rm ee}^{-1}$\,kg$^{-1}$\,d$^{-1}$}}
\newcommand{\um}{\mbox{$\mu$m}}
\newcommand{\sxy}{\mbox{$\sigma_{xy}$}}
\newcommand{\sx}{\mbox{$\sigma_{x}$}}
\newcommand{\smax}{\mbox{$\sigma_{\rm max}$}}
\newcommand{\spix}{\mbox{$\sigma_{\rm pix}$}}
\newcommand{\obo}{\mbox{1$\times$1}}
\newcommand{\obh}{\mbox{1$\times$100}}
\newcommand{\dll}{\mbox{$\Delta LL$}}
\newcommand*\diff{\mathop{}\!\mathrm{d}}
\newcommand*\Diff[1]{\mathop{}\!\mathrm{d^#1}}

\newcommand{\tritium}{\mbox{$^{3}$H}}
\newcommand{\ironfive}{\mbox{$^{55}$Fe}}
\newcommand{\coseven}{\mbox{$^{57}$Co}}
\newcommand{\pbten}{$^{210}$Pb}
\newcommand{\biten}{$^{210}$Bi}

\title{Search for low-mass WIMPs in a 0.6 kg day exposure \\of the DAMIC experiment at SNOLAB}

\author{A.~Aguilar-Arevalo} 
\affiliation{Universidad Nacional Aut{\'o}noma de M{\'e}xico, Mexico City, Mexico} 

\author{D.~Amidei}
\affiliation{Department of Physics, University of Michigan, Ann Arbor, MI, United States}  

\author{X.~Bertou}
\affiliation{Centro At\'omico Bariloche - Instituto Balseiro, CNEA/CONICET, Argentina}

\author{M.~Butner}
\affiliation{Fermi National Accelerator Laboratory, Batavia, IL, United States}
\affiliation{Northern Illinois University, DeKalb, IL, United States }

\author{G.~Cancelo}
\affiliation{Fermi National Accelerator Laboratory, Batavia, IL, United States}

\author{A.~Casta\~{n}eda V\'{a}zquez}
\affiliation{Universidad Nacional Aut{\'o}noma de M{\'e}xico,  Mexico City, Mexico} 

\author{ B.A.~Cervantes Vergara}
\affiliation{Universidad Nacional Aut{\'o}noma de M{\'e}xico, Mexico City, Mexico} 

\author{A.E.~Chavarria}
\affiliation{Kavli Institute for Cosmological Physics and The Enrico Fermi Institute, The University of Chicago, Chicago, IL, United States}

\author{C.R.~Chavez}
\affiliation{Facultad de Ingenier\'{\i}a - Universidad Nacional de Asunci\'on, Paraguay}

\author{J.R.T.~de~Mello~Neto}
\affiliation{Universidade Federal do Rio de Janeiro, Instituto de  F\'{\i}sica, Rio de Janeiro, RJ, Brazil}

\author{ J.C.~D'Olivo}
\affiliation{Universidad Nacional Aut{\'o}noma de M{\'e}xico, Mexico City, Mexico} 

\author{J.~Estrada}
\affiliation{Fermi National Accelerator Laboratory, Batavia, IL, United States}

\author{G.~Fernandez~Moroni}
\affiliation{Fermi National Accelerator Laboratory, Batavia, IL, United States}
\affiliation{Universidad Nacional del Sur, Bahia Blanca, Argentina}

\author{R.~Ga\"ior}
\affiliation{Laboratoire de Physique Nucl\'eaire et de Hautes Energies (LPNHE), Universit\'es Paris 6 et Paris 7, CNRS-IN2P3, Paris, France}

\author{Y.~Guardincerri}
\affiliation{Fermi National Accelerator Laboratory, Batavia, IL, United States}

\author{ K.P.~Hern\'{a}ndez~Torres}
\affiliation{Universidad Nacional Aut{\'o}noma de M{\'e}xico, Mexico City, Mexico} 

\author{F.~Izraelevitch}
\affiliation{Fermi National Accelerator Laboratory, Batavia, IL, United States}

\author{A.~Kavner}
\affiliation{Department of Physics, University of Michigan, Ann Arbor, MI, United States}  

\author{ B.~Kilminster}
\affiliation{Universit{\"a}t Z{\"u}rich Physik Institut, Zurich, Switzerland }

\author{I.~Lawson}
\affiliation{SNOLAB, Lively, ON, Canada }

\author{A.~Letessier-Selvon}
\affiliation{Laboratoire de Physique Nucl\'eaire et de Hautes Energies (LPNHE), Universit\'es Paris 6 et Paris 7, CNRS-IN2P3, Paris, France}

\author{J.~Liao}
\affiliation{Universit{\"a}t Z{\"u}rich Physik Institut, Zurich, Switzerland }

\author{V.B.B.~Mello}
\affiliation{Universidade Federal do Rio de Janeiro, Instituto de  F\'{\i}sica, Rio de Janeiro, RJ, Brazil}

\author{J.~Molina}
\affiliation{Facultad de Ingenier\'{\i}a - Universidad Nacional de Asunci\'on, Paraguay}

\author{J.R.~Pe\~{n}a}
\affiliation{Kavli Institute for Cosmological Physics and The Enrico Fermi Institute, The University of Chicago, Chicago, IL, United States}

\author{P.~Privitera}
\affiliation{Kavli Institute for Cosmological Physics and The Enrico Fermi Institute, The University of Chicago, Chicago, IL, United States}

\author{K.~Ramanathan}
\affiliation{Kavli Institute for Cosmological Physics and The Enrico Fermi Institute, The University of Chicago, Chicago, IL, United States}

\author{Y.~Sarkis}
\affiliation{Universidad Nacional Aut{\'o}noma de M{\'e}xico, Mexico City, Mexico} 

\author{T.~Schwarz}
\affiliation{Department of Physics, University of Michigan, Ann Arbor, MI, United States}  

\author{C.~Sengul}
\affiliation{Kavli Institute for Cosmological Physics and The Enrico Fermi Institute, The University of Chicago, Chicago, IL, United States}

\author{ M.~Settimo}
\affiliation{Laboratoire de Physique Nucl\'eaire et de Hautes Energies (LPNHE), Universit\'es Paris 6 et Paris 7, CNRS-IN2P3, Paris, France}

\author{M.~Sofo~Haro}
\affiliation{Centro At\'omico Bariloche - Instituto Balseiro, CNEA/CONICET, Argentina}

\author{ R.~Thomas}
\affiliation{Kavli Institute for Cosmological Physics and The Enrico Fermi Institute, The University of Chicago, Chicago, IL, United States}

\author{J.~Tiffenberg}
\affiliation{Fermi National Accelerator Laboratory, Batavia, IL, United States}

\author{E.~Tiouchichine}
\affiliation{Centro At\'omico Bariloche - Instituto Balseiro, CNEA/CONICET, Argentina}

\author{ D.~Torres Machado}
\affiliation{Universidade Federal do Rio de Janeiro, Instituto de  F\'{\i}sica, Rio de Janeiro, RJ, Brazil}

\author{F.~Trillaud}
\affiliation{Universidad Nacional Aut{\'o}noma de M{\'e}xico, Mexico City, Mexico} 

\author{X.~You}
\affiliation{Universidade Federal do Rio de Janeiro, Instituto de  F\'{\i}sica, Rio de Janeiro, RJ, Brazil}

\author{J.~Zhou}
\affiliation{Kavli Institute for Cosmological Physics and The Enrico Fermi Institute, The University of Chicago, Chicago, IL, United States}

\collaboration{DAMIC Collaboration}
\noaffiliation

\date{\today}

\begin{abstract}
We present results of a dark matter search performed with a 0.6\,kg\,d exposure of the DAMIC experiment at the SNOLAB underground laboratory.
We measure the energy spectrum of ionization events in the bulk silicon of charge-coupled devices down to a signal of 60\,eV electron equivalent.
The data are consistent with radiogenic backgrounds, and constraints on the spin-independent WIMP-nucleon elastic-scattering cross section are accordingly placed.
A region of parameter space relevant to the potential signal from the CDMS-II Si experiment is excluded using the same target for the first time.
This result obtained with a limited exposure demonstrates the potential to explore the low-mass WIMP region ($<$10\,\gev ) with the upcoming DAMIC100, a 100\,g detector currently being installed in SNOLAB.

\end{abstract}

\maketitle

\section{\label{sec:intro}Introduction}

The DAMIC (dark matter in CCDs) experiment~\cite{Barreto2012264} employs the bulk silicon of scientific-grade charge-coupled devices (CCDs) to detect coherent elastic scattering of weakly interacting massive particles (WIMPs), highly motivated candidates for being the dark matter in the Universe~\cite{Kolb:1990vq, *Griest:2000kj, *Zurek:2013wia}.
By virtue of the low readout noise of the CCDs and the relatively low mass of the silicon nucleus, DAMIC is particularly sensitive to low-mass WIMPs in the Galactic halo with masses in the range 1--20\,\gev , which would induce nuclear recoils of keV-scale energies.

Throughout 2015, dark matter search data were acquired in the SNOLAB laboratory with 8\,Mpix CCDs (2.9\,g each) in dedicated one- to two-month-long periods.
In this paper, we present results from a 0.6\,kg\,d exposure reaching a sensitivity to the spin-independent WIMP-nucleus elastic-scattering cross section $<$\powero{-39}\,cm$^2$ for WIMPs with masses $>$3\,\gev\ and directly probing the signal excess in the CDMS II silicon experiment~\cite{Agnese:2013rvf} with the same nuclear target.

This work establishes the calibration and stable performance of the detector, the understanding of backgrounds, and the analysis techniques necessary for the full deployment of the eighteen 16\,Mpix CCDs (5.8\,g each) of DAMIC100.

\section{\label{sec:detector}DAMIC Experiment at SNOLAB}

\begin{figure}
\centering
\includegraphics[width=0.4\textwidth]{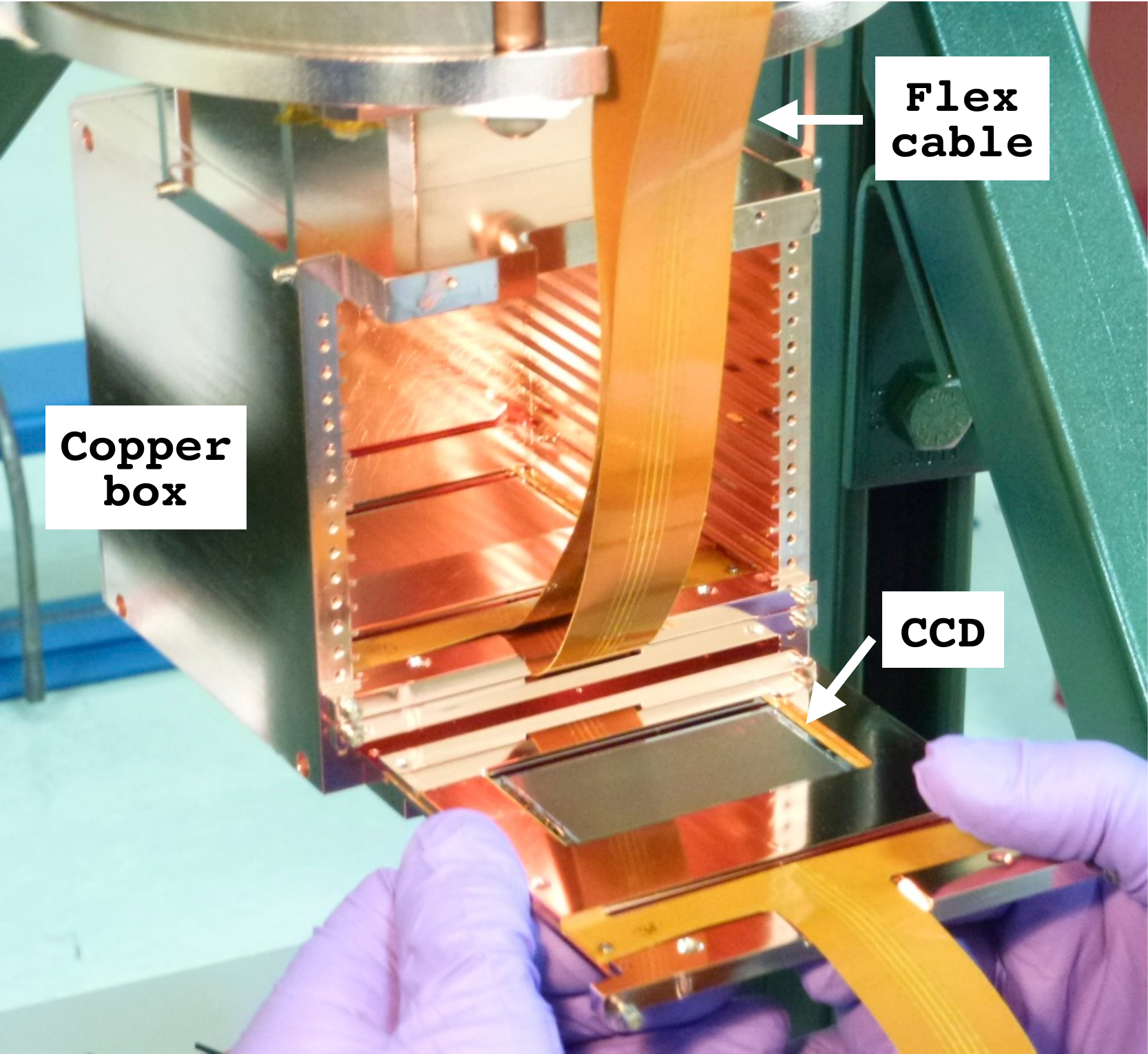}
\caption{Copper module holding an 8\,Mpix CCD being installed in the low-radioactivity copper box. Two other modules have already been installed and can be partially seen at the bottom of the box. The flex cables that carry the CCD signals are also visible.} \label{fig:det}
\end{figure}

The DAMIC CCDs are packaged in a copper module including a silicon support frame for the CCD and a low-radioactivity flex cable to carry the signals that drive and read the device (Fig.~\ref{fig:det}).
The modules are inserted in slots of a copper box that is cooled to $\sim$120\,K inside a copper vacuum vessel ($\sim$\powero{-6}\,mbar). 
The box is shielded on all sides by lead to attenuate external $\gamma$ rays. A 18-cm-thick lead shield is suspended immediately above the box inside the vacuum vessel, and a lead castle of 21\,cm thickness shields the copper vessel from all other sides. 
The innermost inch of lead comes from an ancient Spanish galleon and has negligible \pbten\ content, strongly suppressing the background from bremsstrahlung $\gamma$'s produced by \biten\ decays in the outer lead shield.
A 42-cm-thick polyethylene shield is used to moderate and absorb environmental neutrons.
The overburden of the laboratory site (6010\,m water equivalent) strongly suppresses the cosmic muon flux to a negligible level of $<$0.27\,m$^{-2}$\,d$^{-1}$~\cite{snolabuh}.
Details of the DAMIC infrastructure at SNOLAB can be found in Ref.~\cite{Aguilar-Arevalo:2015lvd}.

\begin{figure}
\centering
\includegraphics[width=0.48\textwidth]{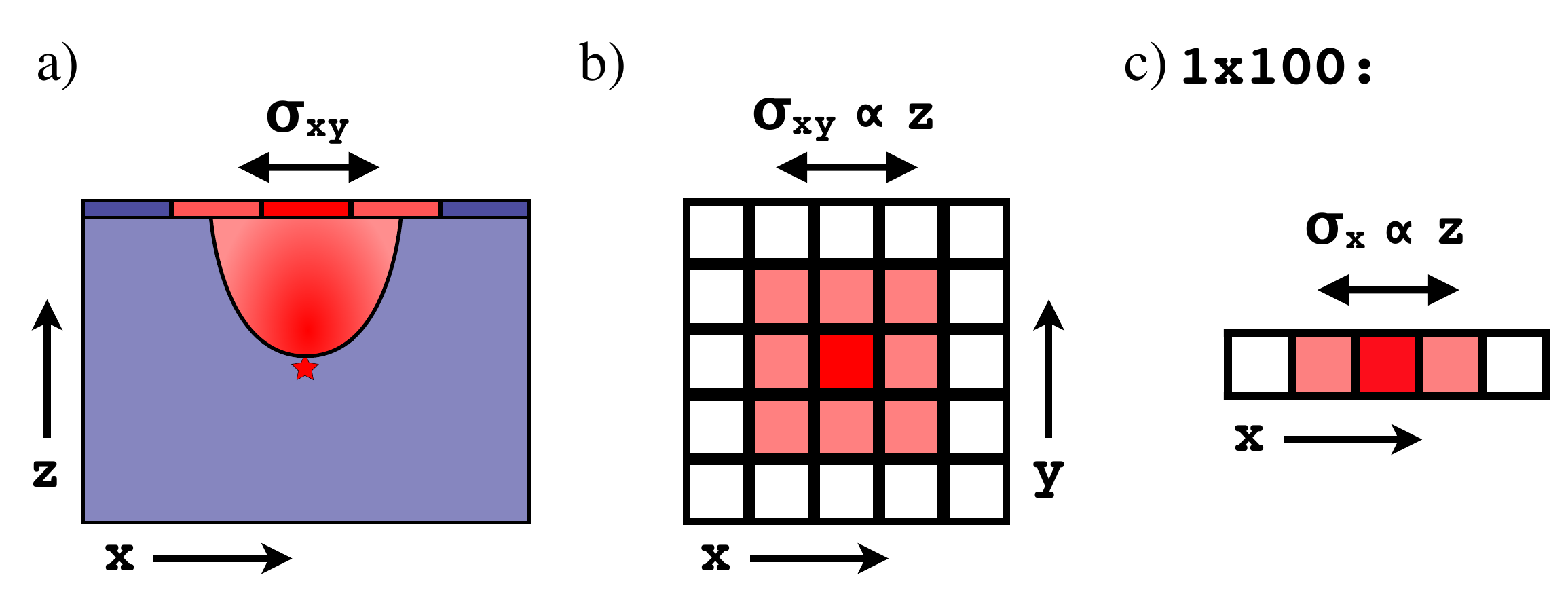}
\caption{a) Cross-sectional representation of the charge produced by a point-like ionization event (star) in the CCD bulk as it is drifted to the pixel array. b) The $x$-$y$ distribution of charge on the pixel array following the ionization event. Because of diffusion, the charge is collected in multiple pixels, with the lateral spread (\sxy) being positively correlated with the depth ($z$ coordinate) of the interaction. When the CCD is read out in the \obo\ configuration, this is the pattern observed in the image. c) In the \obh\ mode, the CCD is read out in column segments 100 pixels tall, collapsing the pixel contents along the $y$ axis, leading to a one-dimensional pattern with the charge spread out over fewer pixels. The one-dimensional lateral spread (\sx ) is positively correlated to the depth of the interaction.} \label{fig:pix}
\end{figure}

The DAMIC CCDs were developed at Lawrence Berkeley National Laboratory MicroSystems Lab~\cite{1185186}, starting from an existing  design for the Dark Energy Survey camera~\cite{Mclean:2012pka}.
They feature a three-phase polysilicon gate structure with a buried $p$ channel.
The pixel size is 15$\times$15\,$\mu$m$^{2}$ and the bulk of the device is high-resistivity (10--20 k$\Omega$\,cm) $n$-type silicon with a thickness of 675\,\um.
The high resistivity of the silicon allows for a low donor density in the substrate ($\sim$\powero{11}\,cm\up{-3}), which leads to fully depleted operation at a substrate bias of 40\,V.
Ionization charge produced in the bulk is drifted along the direction of the electric field ($z$ axis).
The holes (charge carriers) are collected and held near the $p$-$n$ junction, less than 1\,\um\ below the gates.
Because of thermal motion, the ionized charge diffuses transversely with respect to the electric field direction as it is drifted  [Fig.~\ref{fig:pix}(a)], with a spatial variance ($\sigma_{x}^2$$=$$\sigma_{y}^2$$=$$\sigma_{xy}^2$) that is proportional to the carrier transit time.
Hence, there is a positive correlation between the lateral diffusion (\sxy ) of the collected charge on the pixel array and the depth of the interaction ($z$).
The maximum observed lateral spread ($\sim$20\,\um) occurs for ionization events on the back surface of the device for which $\sim$25 pixels collect 95\% of the generated charge carriers.

\section{\label{sec:readout}CCD Readout}

The ionized charge is collected and held at the gates throughout hour- to day-long image exposures until the device is read out.
During readout, the charge is transferred in the $y$ direction from pixel to pixel along each column by appropriate clocking of the three-phase gates (``parallel clocks''), while higher frequency clocks (``serial clocks'') move the charge of the last row (the ``serial register") in the $x$ direction to the CCD's output node, where the charge is measured by a correlated double-sampling circuit~\cite{janesick2001scientific}.
The inefficiency of charge transfer from pixel to pixel~\cite{janesick2001scientific} is as low as \powero{-6}~\cite{1185186}, and the readout noise for the charge collected in a pixel is $\sim$2\,$e^-$ (Sec.~\ref{sec:dataset}). 
The image is reconstructed from the order in which the pixels are read out, and contains a two-dimensional stacked history (projected on the $x$-$y$ plane) of all particle interactions throughout the exposure.
Fig.~\ref{fig:pix}(b) shows the pattern observed on the $x$-$y$ plane from an ionization event in the CCD bulk.
The number of pixels above a given threshold due to noise fluctuations is proportional to the total number of pixels read out.
Therefore, it is advantageous for rare-event searches to take the longest possible exposures.
Given the small dark current of the CCDs ($<$\powero{-3}\,$e^-$pix$^{-1}$day$^{-1}$ at the operating temperature of $\sim$120\,K), exposures up to several days can be taken without introducing additional noise.

With appropriate clocking, the charge of multiple adjacent pixels can be added in the output node before the charge measurement is performed. 
DAMIC data have been acquired so far with two different readout configurations: \obo\ and \obh .
The first configuration is the standard CCD readout, where the charge collected by each pixel is read out individually, offering maximum spatial resolution.
In the latter configuration, 100 rows are transferred into the serial register before the charge is clocked in the $x$ direction, and each column segment is read out individually.
As the total charge of an ionization event is distributed over a smaller number of charge measurements, there is a smaller contribution from the readout noise.
As a consequence, the energy resolution and the energy threshold for ionization events distributed over multiple pixels is improved.
However, the spatial resolution in the $y$ coordinate is lost, with \sx\ still positively correlated to the depth of the interaction [Fig.~\ref{fig:pix}(c)].
DAMIC CCDs are read out with an integration time for the correlated double sampling of 40\,$\mu$s, which leads to an image readout time of 840\,sec (20\,sec) in the \obo~(\obh ) mode.

DAMIC CCDs feature an output node at each end of the serial register. As described above, all the charge collected by the CCD pixel array is read out through one of these output nodes.
No charge is deposited in the second output node, which is also read out and offers a measurement of zero charge, i.e., of noise.
Since the readout of the two output nodes is synchronized by the clocking, the noise measurement by the second output node allows the identification and suppression of the correlated electronic noise of the detector's readout chain (Sec.~\ref{sec:dataset}).

\section{\label{sec:calib}Energy and depth response of a DAMIC CCD}

\subsection{\label{sec:energy}Energy}
The output of a CCD readout chain is recorded in analog-to-digital converter units (ADU) proportional to the number of charge carriers placed in the CCD's output node.
The signals produced by recoiling electrons, which lose their energy through ionization, are proportional to the generated number of charge carriers, with an average of one electron-hole pair produced for every 3.77\,eV of deposited energy~\cite{4326950}.
Thus, we define the electron-equivalent energy scale (in units of \eve ) relative to the ionization produced by recoiling electrons from the photoabsorption of x rays of known energy.

\begin{figure}
\centering
\includegraphics[width=0.48\textwidth]{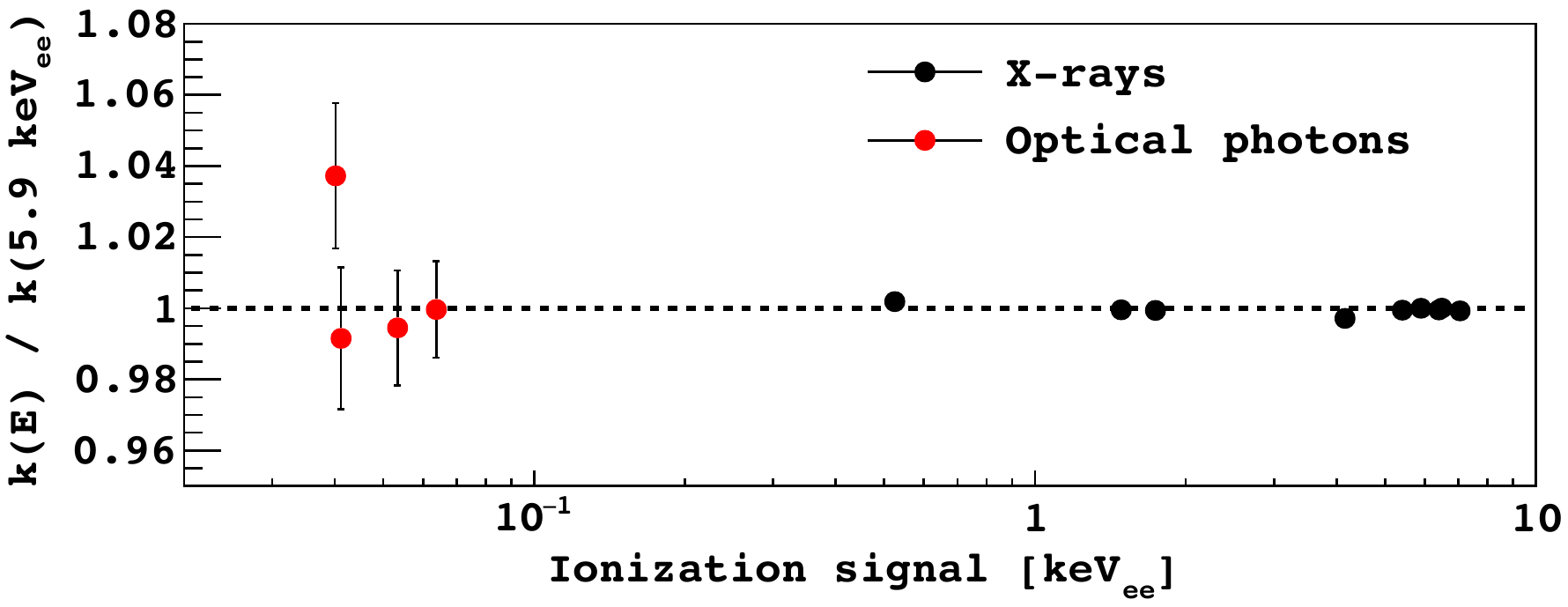}
\caption{Linear constant $k$ relating the CCD output signal to the ionization generated in the substrate. Values are given relative to $k$ measured at 5.9\,k\eve.  Calibrations at high energies were performed with x rays, while the lowest energy points were obtained using optical photons, as outlined in the text.
The linearity of the CCD energy response is demonstrated down to 40\,\eve .} \label{fig:elin}
\end{figure}

Calibrations were performed by illuminating the CCD with fluorescence x rays from O, Al, Si, Cr, Mn, and Fe.
Fig.~\ref{fig:elin} summarizes the measurement of the linear calibration constant, $k$ (in units of ADU/\eve ), at different energies, which demonstrates the linear response of the CCD to electron recoils.
From x-ray data we also estimated the intrinsic fluctuations in the number of charge carriers produced.
The measured resolution of 54\,\eve\ at 5.9\,k\eve\ corresponds to a Fano factor~\cite{PhysRev.72.26, *doi:10.1117/12.948704} of 0.133$\pm$0.005.

To demonstrate the linearity of the CCD output to lower-energy signals, we used optical photons from a red light-emitting diode (LED) installed inside the DAMIC copper vessel, which produce a single electron-hole pair by photoelectric absorption.
Several CCD images were read out, each exposed to light for 20\,sec.
For a given pixel, the number of charge carriers detected in the images follows a Poisson distribution.
The mean ($\mu_l$) and variance ($\sigma_l^2$) of the increase in the pixel ADU induced by the LED exposure are then related to the calibration constant ($k$) by
\begin{equation}
k = \frac{1}{3.77\mbox{\,\eve }}\frac{\sigma_l^2}{\mu_l}.
\label{eq:poisson}
\end{equation}
We employed Eq.~(\ref{eq:poisson}) to estimate the calibration constant at very low light levels, when only a few of charge carriers are collected by a pixel.
These results are included in Fig. 3 and demonstrate a CCD response linear within 5\% down to 40\,\eve .

A recoiling silicon nucleus following a WIMP interaction in the CCD bulk will deposit only a fraction of its energy through ionization, producing a significantly smaller signal than a recoiling electron of the same energy. The nuclear recoil ionization efficiency, which relates the ionization signal in the detector (in units of \eve ) to the kinetic energy of the recoiling nucleus (in units of \evr ), must be known to properly interpret the measured ionization spectrum in terms of WIMP-induced recoils. Until recently, measurements of the nuclear recoil ionization efficiency in silicon were available only down to $\sim$3\,k\evr~\cite{PhysRevD.42.3211}, and a theoretical model from Lindhard {\it et al.}~\cite{Lindhard:1963vo, *ziegler1985stopping} was usually employed to extrapolate to lower energies. We adopt new results~\cite{antonella1, Chavarria:2016xsi} that extend the measured nuclear recoil ionization efficiency down to $\sim$0.7\,k\evr ,  covering most of the energy range relevant for low-mass WIMP searches. Measurements in~\cite{antonella1} employ a silicon drift detector exposed to a fast-neutron beam at the Tandem Van de Graaff facility of the University of Notre Dame to provide accurate results in the 2--20\,k\evr\ energy range. For the calibration at lower energies~\cite{Chavarria:2016xsi}, nuclear recoils were induced in a DAMIC CCD by low-energy neutrons from a $^{124}$Sb-$^9$Be photoneutron source, and their ionization signal was measured down to 60\,\eve . A linear extrapolation of these results is used for the nuclear recoil ionization efficiency below 60\,\eve , resulting in no ionization signal for nuclear recoils below 0.3$\pm$0.1\,k\evr  .

\subsection{\label{sec:depth}Depth}

The relationship between \sxy\ and $z$ of an interaction can be analytically solved in one dimension given the electric field profile within the CCD substrate and the fact that the lateral variance of the charge carriers ($\sigma_{xy}^2$) due to diffusion is proportional to the transit time from the interaction point to the CCD pixel array~\cite{1185186}.
The resulting relation is
\begin{equation}
\sigma_{xy}^2 = -A\ln|1-bz|.
\label{eq:xyzrel}
\end{equation}
The constants $A$ and $b$ are related to the physical properties and the operating parameters of the device and are given by
\begin{eqnarray*} 
&A = \frac{\epsilon}{\rho_n} \frac{2k_BT}{e}, \\
&b = \left( \frac{\epsilon}{\rho_n} \frac{V_b}{z_D} + \frac{z_D}{2} \right)^{-1},
\label{eq:abl}
\end{eqnarray*}
where $\epsilon$ is the permittivity of silicon, $\rho_n$ is the donor charge density in the substrate, $k_B$ is Boltzmann's constant, $T$ is the operating temperature, $e$ is the electron's charge, $V_b$ is the bias applied across the substrate, and $z_D$ is the thickness of the device.

\begin{figure}
\centering
\includegraphics[width=0.48\textwidth]{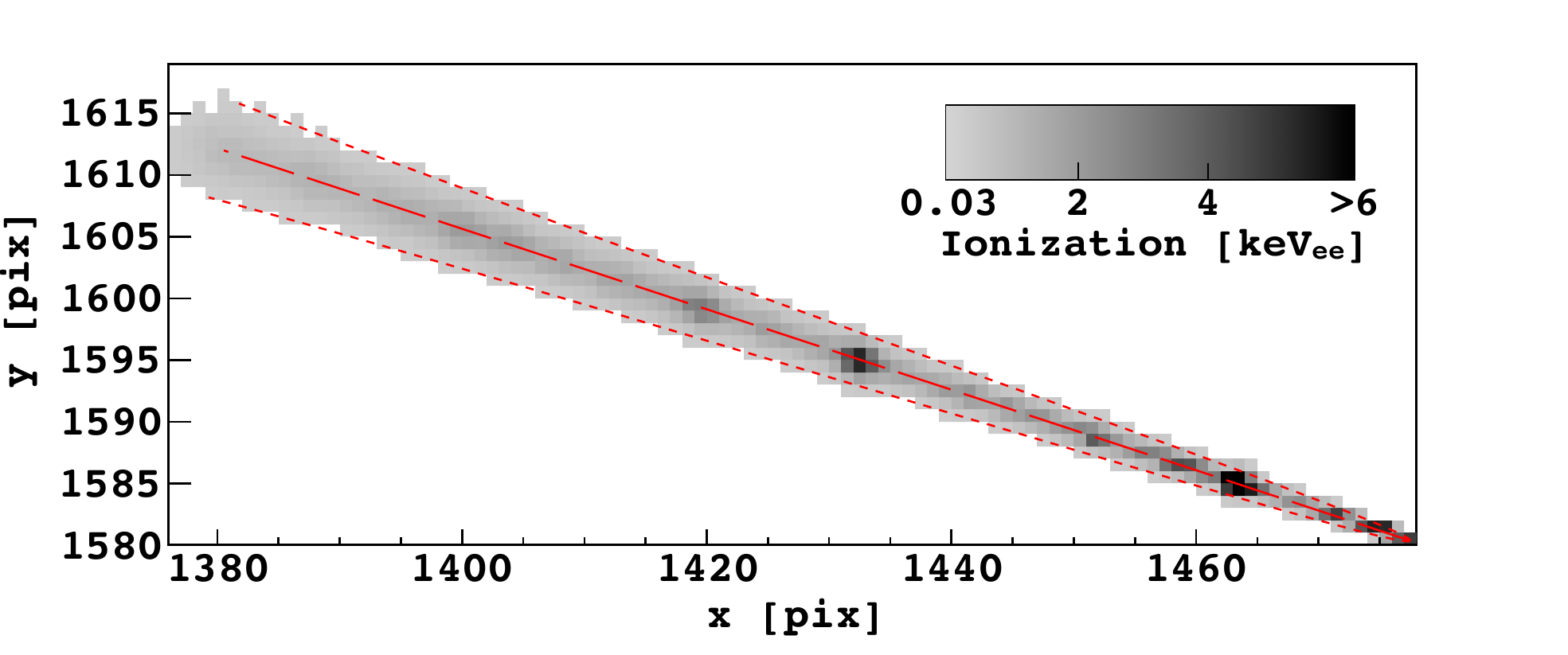}
\caption{A MIP observed in cosmic ray background data acquired on the surface. Only pixels whose values are above the noise in the image are colored. The large area of diffusion on the top left corner of the image is where the MIP crosses the back of the CCD. Conversely, the narrow end on the bottom right corner is where the MIP crosses the front of the device. The reconstructed track is shown by the long-dashed line. The short-dashed lines show the 3\,$\sigma$ band of the charge distribution according to the best-fit diffusion model.} \label{fig:diff}
\end{figure}

In practice, it is most accurate to measure the parameters $A$ and $b$ directly from data.
This was done using cosmic ray background data acquired on the surface, by fitting the width of minimum ionizing particles (MIPs) that penetrate the CCD as a function of depth.
These events are identified as straight tracks with a relatively constant energy deposition per unit length consistent with the stopping power of a MIP.
As MIP tracks follow a straight line, the depth can be calculated unambiguously from the path length on the $x$-$y$ plane.
Fig.~\ref{fig:diff} shows a MIP in a CCD operated at the nominal temperature and substrate bias used in SNOLAB.
Characteristic bursts of charge (darker spots) along the track correspond to the emission of $\delta$ rays.
The best-fit parameters to the diffusion model [Eq.~(\ref{eq:xyzrel})] are $A$$=$215$\pm$15\,\um$^2$ and $b$$=$\powert{-3}{(1.3$\pm$0.1)}\,\um$^{-1}$, which correspond to a maximum diffusion at the back of the device of \smax$=$(21$\pm$1)\,\um\,$\approx$\,1.4\,pix.
The accuracy of this calibration has been validated by studying the diffusion of x-ray events that interact near the surfaces on the back and the front of the  CCD~\cite{Chavarria201521} and from $\gamma$-ray data, which provide ionization events uniformly distributed in the bulk of the device.

By studying the energy loss of $\beta$'s from an external \tritium\ source, we find that the CCD has a $\sim$2\,\um\ dead layer on its front and back surfaces, as expected from the fabrication process of the device~\cite{1185186}. There is no evidence for regions of partial or incomplete charge collection that may hinder the CCD energy response.

\section{\label{sec:dataset}Data sets and image processing}

\begin{table*}[t!]
\caption{\label{tab:dataruns}Summary of the data runs used for the dark matter search.}
\begin{ruledtabular}
\begin{tabular}{cccccc}
Start date & End date & Acquisition mode & No. of CCDs & No. of exposures & Total exposure (kg\,d)\\
\hline
2014/12/12 & 2015/02/17 & \obo\ & 2 & 225 & 0.235\\
2015/07/06 & 2015/07/20 & \obo\ & 3 & 18 & 0.056\\
2015/10/28 & 2015/12/05 & \obo\ & 3 & 29 & 0.091\\
2015/02/01 & 2015/02/18 & \obh\ & 2 & 65 & 0.040\\
2015/04/21 & 2015/05/04 & \obh\ & 2 & 104 & 0.065\\
2015/07/06 & 2015/07/20 & \obh\ & 3 & 18 & 0.017\\ 
2015/10/28 & 2015/12/05 & \obh\ & 2 & 44 & 0.082\\
\end{tabular}
\end{ruledtabular}
\end{table*}

The DAMIC setup at SNOLAB was devoted to background studies throughout the years 2013--2015, with more than ten installations involving changes to the external shielding and CCD packaging and different materials being placed inside the copper box for screening purposes.
During 2015, data were acquired intermittently in both \obo\ and \obh\ acquisition modes with two or three 8\,Mpix, 675\,\um -thick CCDs (2.9 g each).
Table~\ref{tab:dataruns} summarizes the dark matter search data runs including the number of CCDs  and images, and the total exposure after the mask and image selection procedures discussed below.

The energy and diffusion responses of all CCDs were calibrated with x rays and cosmic rays on the surface before deployment.
At SNOLAB, a fluorescence copper line (8\,keV) induced by radioactive particle interactions in the copper surrounding the CCDs was used to confirm the calibrated energy scale.
The value of \smax\ was also monitored to validate the depth response calibrated on the surface.
The radiogenic background rate measured below 10\,k\eve\ decreased with time thanks to the continuous improvements in the radio purity of the setup, with an average event rate throughout the data runs of $\sim$30\,\dru .

Images were taken with exposures of either $10^4$ or 3$\times$$10^{4}$ sec, immediately followed by the acquisition of a ``blank"  image whose exposure is only the readout time.
Because of the $<$5\% ($<$0.1\%) probability of a physical event occurring during \obo\ (\obh ) readout, most blanks contain only the image noise.

Each image was processed as follows.
First, a pedestal was subtracted from each pixel value, estimated as the median of the pixel values of the column to which the pixel belongs.
Correlated noise results in a simultaneous shift of the pedestal value at the two output nodes of the serial register.
This shift was estimated by fitting a linear relation to the values read out by both output nodes for pixels along a row (Sec.~\ref{sec:readout}) and was then subtracted.

For each data run (Table~\ref{tab:dataruns}) we calculated the median and median absolute deviation (MAD) of every pixel over all images in the run. These quantities are used to construct a ``mask," which excludes pixels which either deviate more than 3\,MAD from the median in at least 50\% of the images or have a median or MAD that is an outlier when compared to the distributions of these variables for all pixels.

\begin{figure}[t!]
\centering
\includegraphics[width=0.48\textwidth]{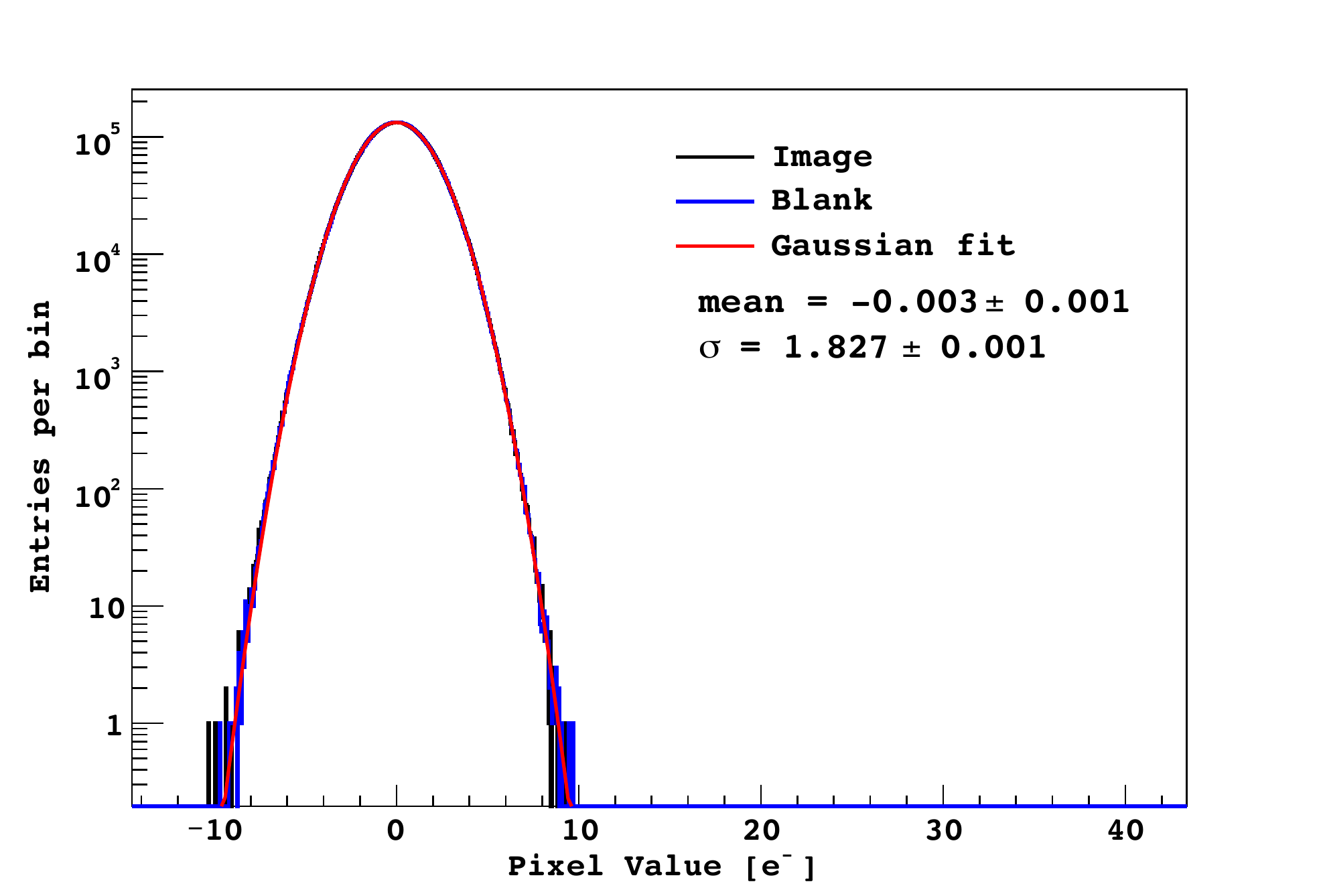}
\caption{Example of the pixel value distribution after image processing in one 30\,ks exposure (black) and its corresponding blank (blue) acquired in December 2014. The noise in the image is fitted to \spix$=$1.8\,$e^-$.} \label{fig:pixdist}
\end{figure}

Figure~\ref{fig:pixdist} shows an example of the distribution of pixel values after pedestal and correlated noise subtraction for a single 30\,ks exposure compared to its corresponding blank.
The blank distribution is accurately described by a Gaussian centered at zero with pixel noise \spix$=$1.8\,$e^-$$\approx$\,7\,\eve .
The 30\,ks exposure presents a statistically consistent white noise distribution, allowing for the identification of a pixel that has collected $>$10\,$e^-$$\approx$\,40\,\eve\ from ionization.

The consistency between each image and its blank was checked by comparing their noise distributions.
Images for which there is a significant discrepancy between the two distributions or for which the distributions deviate from white noise were excluded from the analysis.
This includes some CCDs in runs acquired between February and August 2015, where the pixel noise was relatively high ($\sim$2.2\,$e^-$).
During this period, the polyethylene shield was partially open, and a small amount of light leaked into the vessel, producing an increased background charge in some of the CCDs.

\section{Event reconstruction and selection\label{sec:erecsel}}

The dark matter search was limited to events with energies $<$10\,k\eve , for which the track length of the ionizing particle is much smaller than the pixel size, and the energy deposition may be considered pointlike. 
Thus, we masked all high-energy ionization events identified as clusters of contiguous pixels with signal larger than 4\,\spix\ whose total collected charge amounts to $\ge$10\,k\eve.
In addition, pixels that were less than four pixels away from the cluster or less than 50 pixels to the left of the cluster (i.e., within 50 subsequent pixel readouts) were masked in the \obo\ data set.
Pixels that were less than 200 pixels to the left of the cluster were masked in the \obh\ data set.
This condition rejected pixels with stray charge due to CCD charge transfer inefficiencies, which may happen when a high-energy interaction results in a large number of charge carriers in the serial register.
The average fraction of masked pixels in an image, including those discarded by the criteria outlined in Sec.~\ref{sec:dataset}, was 1\% (8\%) in the \obo~(\obh ) data set.

A likelihood clustering algorithm based on a (11$\times$11)-pixel window moving over the unmasked regions was then applied to the \obo\ data set.
For every position of the window, we computed i)~the likelihood $\mathcal{L}_{n}$ that the pixel values in the window are described by white noise and ii)~the likelihood $\mathcal{L}_{G}$ that the pixel values in the window are described by a two-dimensional Gaussian function on top of white noise, where the expected value of pixel $(i,j)$ is
\begin{equation*}
f_{G}(i,j) = I \int_{i-\frac{1}{2}}^{i+\frac{1}{2}} \int_{j-\frac{1}{2}}^{j+\frac{1}{2}} \mbox{Gaus} \left(x,y | \mu_{x},\mu_{y}, \sigma_{x}, \sigma_{y}\right) \diff x \diff y
\end{equation*}
with the Gaussian parameters fixed: $\mu_{x}$ and $\mu_{y}$ to the values of the coordinates of the center of the window, the standard deviations  \sx$=$$\sigma_{y}$=\sxy\ to a value of one pixel, and the integral $I$ to the sum of pixel values in the window.
We considered a candidate cluster to be present in the search window when $-\ln[\mathcal{L}_{G}/\mathcal{L}_{n}]$$<$$-4$  (i.e., there is a significant preference for the Gaussian hypothesis).
The search window was then moved around to find the local minimum of this quantity, where the window position was fixed and a fit was performed, leaving $I$, $\mu_{x}$, $\mu_{y}$, and \sxy\ as free parameters to maximize the value of  $\mathcal{L}_{G}$.
Our best estimates for the number of collected charge carriers, the $x$-$y$ position of the cluster and the lateral spread of the charge were obtained from the fitted parameters as $N_{e}$$=$$I/(k\times 3.77\mbox{\,\eve })$, $\mu_{x}$, $\mu_{y}$, and \sxy , respectively.
The cluster energy ($E$) was then derived from $N_{e}$ (Sec.~\ref{sec:calib}).
The test statistic
\begin{equation*}
\Delta LL=-\ln\left[\frac{\rm{max}(\mathcal{L}_{G})}{\mathcal{L}_{n}}\right]
\end{equation*}
was also calculated, which gives the significance of a cluster to originate from an ionization event and not from white noise. 

In the \obh\ acquisition mode, the clustering procedure is very similar, except that it is performed in one dimension along rows of the image.
The fitting function $f_G$ is reduced to a one-dimensional Gaussian with  $\mu_{x}$ and \sx\ as free parameters.
The interpretations of the best fit values are analogous.

\begin{figure}[t!]
\centering
\includegraphics[width=0.48\textwidth]{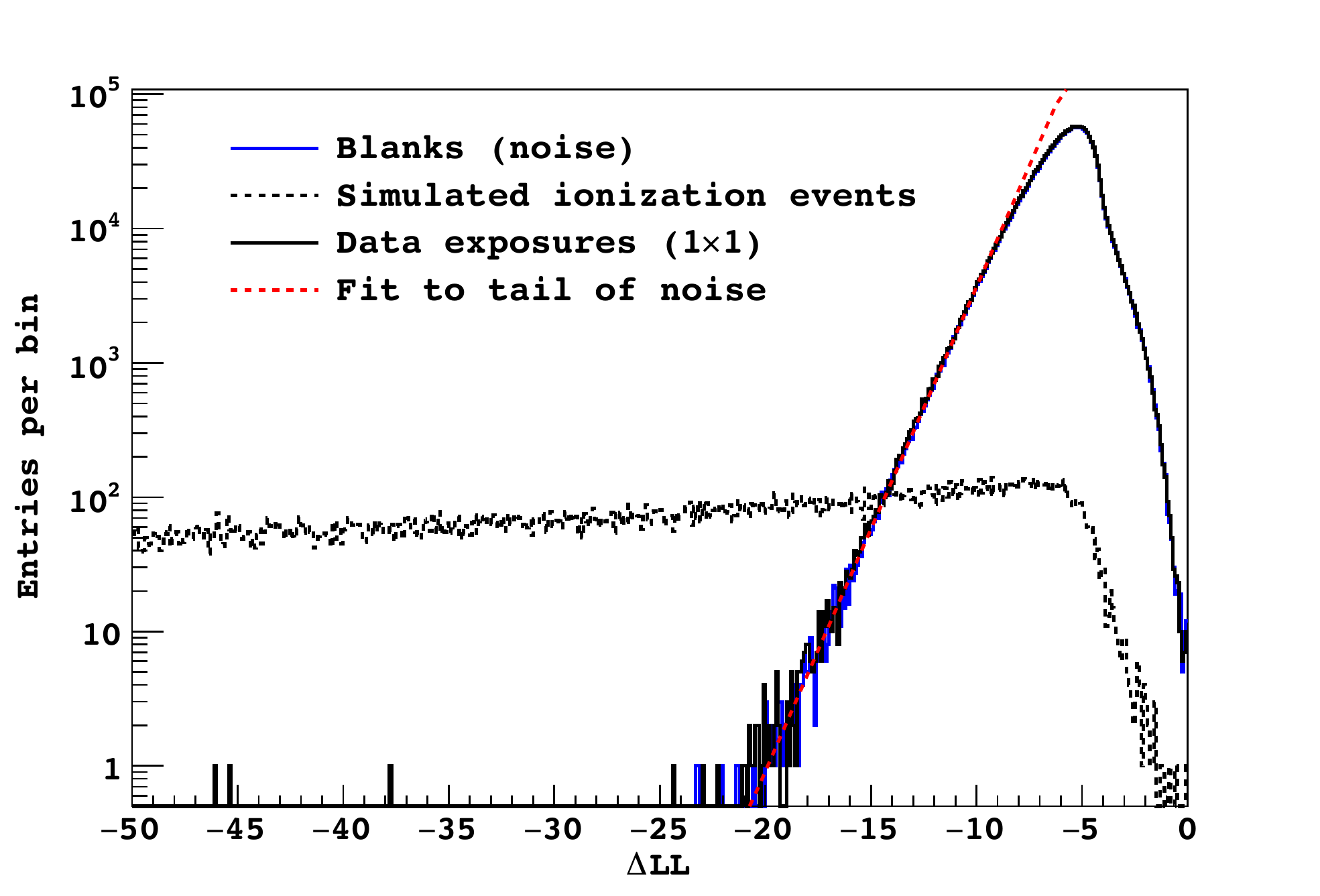}
\caption{ \dll\ distributions for all clusters in the \obo\ data set. The blue line shows the distribution for clusters in the blanks, which are representative of the contribution from readout noise to the data set. The black dashed line presents the expected distribution (from simulation) of ionization events that occur uniformly in the CCD bulk, assuming a constant (flat) energy spectrum. The black line shows the distribution for all clusters in the \obo\ data set. The dashed red line is the fit done to the tail of the noise distribution to determine the selection used to reject readout noise. The fit is statistically consistent with the tail of the distribution.} \label{fig:dll}
\end{figure}

Fig.~\ref{fig:dll} shows the \dll\ distribution of all clusters in the \obo\ data set and their corresponding blanks.
Clusters due purely to noise have the same \dll\ distribution in data images and blanks, with an exponentially decreasing tail at low \dll\ values, as expected for white noise.
They were rejected by requiring \dll$<$$-28$ ($-25$) for the \obo~(\obh ) data set.
From an exponential fit to the tail of the \dll\ distribution, we estimate that  $<$0.01 background noise clusters are left in each data set after this selection.

\begin{figure*}[t!]
\centering
\includegraphics[width=0.7\textwidth]{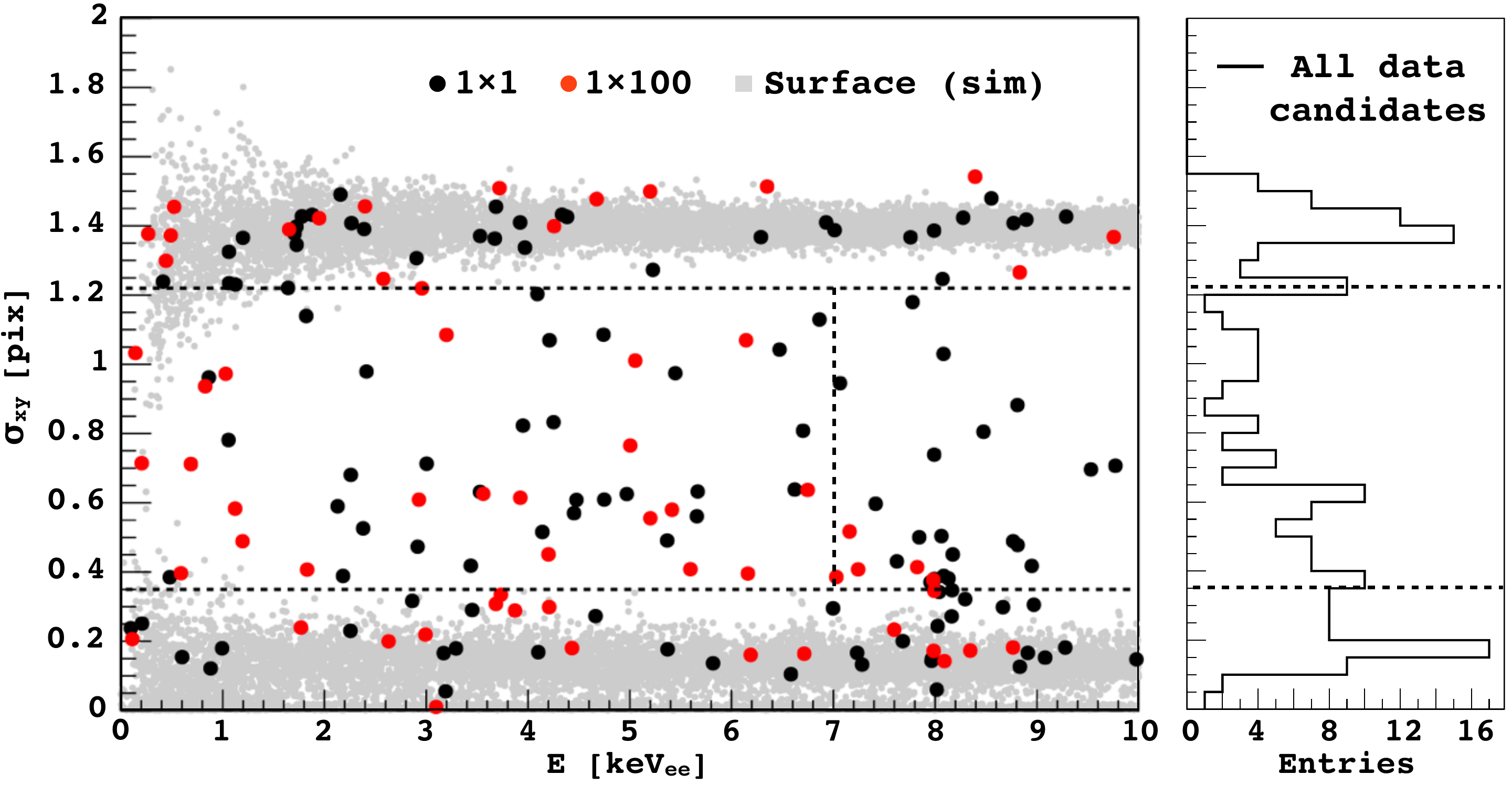}
\caption{Lateral spread (\sxy ) versus measured energy ($E$) of the clusters that pass the selection criteria outlined in Sec.~\ref{sec:erecsel}. Black (red) markers correspond to candidates in the \obo~(\obh ) data set. Gray markers show the simulated distribution of energy deposits near the front and back surfaces of the device. The projection on the \sxy\ axis of the identified clusters is shown on the right. The horizontal dashed lines represent the fiducial selection described in Sec.~\ref{sec:fidu}, while the vertical dashed line shows the upper bound of the WIMP search energy range.} \label{fig:sige}
\end{figure*}

 In the selected sample, we noticed some recurring events in the same spatial position in the CCDs.
 These events arise from small defects in the silicon lattice that produce an increased level of dark current at a specific spatial position.
As these events are very faint, they were missed by the masking criteria outlined in Sec.~\ref{sec:dataset}.
We removed them from the final candidates with a negligible impact on the acceptance, as the probability of two uncorrelated events occurring in the same pixel is $\ll$0.1\%.
Likewise, we excluded clusters that were less than 300\,\um\ on the $x$-$y$ plane from any other cluster in the same image.
These spatially correlated clusters are likely to be produced by radiation following radioactive decay and unlikely to arise from WIMP interactions.
Their exclusion also has a negligible impact on the acceptance.
After the application of these criteria, 122 (62) final candidate clusters remain in the \obo\ (\obh) data sets.
Fig.~\ref{fig:sige} shows the lateral spread versus energy distribution of the candidates.

We estimated the performance of the reconstruction algorithm for WIMP-like events by Monte Carlo simulations.
Pointlike interactions with deposited energy in the range of interest were simulated following a uniform spatial distribution in the CCD bulk.
For each simulated event, the charge distribution on the pixel array was derived according to the diffusion model (Sec.~\ref{sec:depth}).
We then added 2000 (200) simulated events on each of the acquired \obo~(\obh ) raw data images to include a realistic noise distribution.
The full data processing chain was run on each image, including the signal identification and likelihood clustering.
Fig.~\ref{fig:dll}  shows the \dll\ distribution of the simulated events selected in the \obo\ data set (dashed black).
We found no bias within 1\% in the reconstructed energy of simulated events with $E$$>$100\,\eve .
A small overestimation  may be present at lower energies, to at most 5\% at the 60\,\eve\ threshold.
With this sample of simulated events, we also estimated the resolution in the ionization signal to be $\sigma_0$$=$37\,\eve\ (30\,\eve ) in the \obo~(\obh ) data set.
Thus, the energy response of the detector can be modeled with a resolution $\sigma_{\rm res}^2$$=$$\sigma_0^2$$+$$(3.77{\rm \,eV_{ee}})FE$, where $F$ is the Fano factor.

The event selection efficiency was estimated from the fraction of simulated events that pass the event selection criteria. For events uniformly distributed in the CCD bulk, the selection efficiency was found to increase from 9\%~(25\%) at 75\,\eve~(60\,\eve ) to $\sim$100\% at 400\,\eve~(150\,\eve ) in the \obo~(\obh ) data.

The better energy resolution and higher selection efficiency of lower-energy events in the \obh\ data set are due to the improved signal to noise of events originating deeper in the bulk of the device, which experience significant  lateral charge diffusion.

\section{\label{sec:fidu}Rejection of surface events}

The selection criteria presented in Sec.~\ref{sec:erecsel} were implemented to distinguish events due to ionization by particle interactions from electronic noise.
High-energy photons that Compton scatter in the bulk of the device produce background ionization events with a uniform spatial distribution because the scattering length is always much greater than the thickness of the CCD.
Hence, ionization events from Compton scattering are only distinguishable from WIMP interactions through their energy spectrum.
Nuclear recoils from WIMP interactions would produce a characteristic spectrum that decreases exponentially with increasing energy, while the Compton scattering spectrum is almost flat throughout the WIMP search energy region.

Another background comes from low-energy electrons and photons radiated by surfaces surrounding the CCDs, and from electrons produced in the silicon that exit the device after depositing only a small fraction of their energy.
These events occur tens of \um\ or less from the surface of the CCDs and can be rejected by appropriate requirements on the depth of the interaction.
We selected events with 0.35$<$\sxy$<$1.22, corresponding to interactions that occur more than 90\,\um\ and 75\,\um\ from the front and back surfaces, respectively, which left 51 (28) candidates in the \obo~(\obh ) data set.
The dashed lines in Fig.~\ref{fig:sige} represent this fiducial region.
The group of events at 8\,k\eve\ corresponds to Cu fluorescence x rays from radioactive background interactions in the copper surrounding the CCDs.
Because of the relatively long x-ray absorption length at this energy (65\,\um ), some of the events leak into the fiducial region.
We, thus, restricted the WIMP search  to clusters with energies $<$7\,k\eve .
The selection efficiency was estimated by simulation to be $\sim$75\% for events uniformly distributed in the CCD bulk.

\begin{figure}[t!]
\centering
\includegraphics[width=0.48\textwidth]{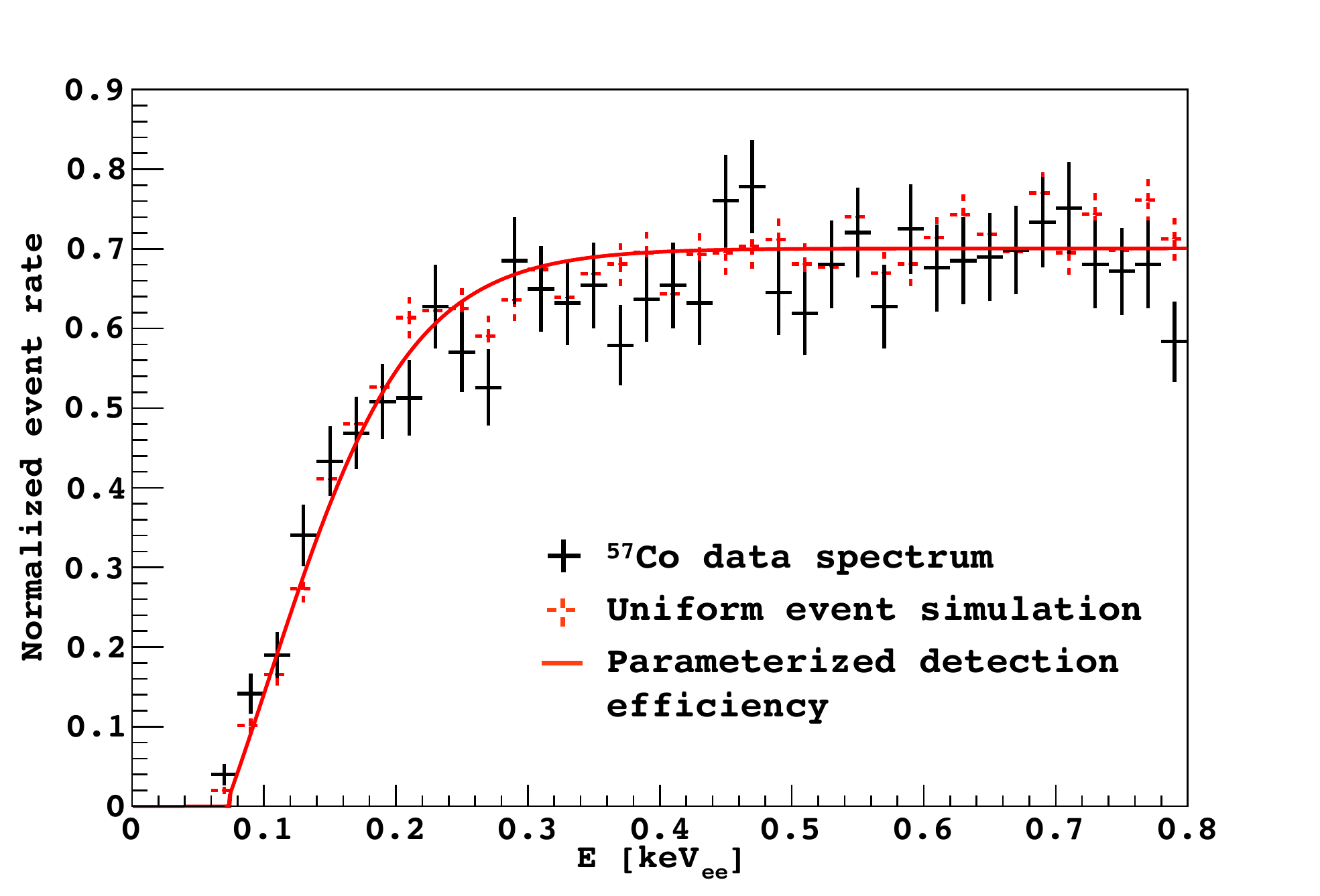}
\caption{Spectrum from \coseven\ source calibration in the laboratory after event selection to remove readout noise and surface events, as performed in the WIMP search. The event rate has been normalized to the absolute rate expected in the energy interval 0.5--1.5\,k\eve . The spectrum is taken as a direct measurement of the detection efficiency because the Compton scattering spectrum at these low energies is approximately constant. The simulated detection efficiency, including the fit with the functional form used for the WIMP search analysis, is shown.} \label{fig:co57}
\end{figure}

To validate our procedure to estimate the detection efficiency, we performed an analogous analysis using \coseven\ $\gamma$-ray calibration data acquired in the laboratory with a 500-\um-thick, 8\,Mpix DAMIC CCD operated with the same \obo\ settings as in SNOLAB. We applied the same image processing, clustering, and event selection criteria outlined in Sec.~\ref{sec:erecsel} and selected a range in \sxy\ to reject events less than 90\,\um\ and 75\,\um\ from the front and back surfaces, respectively. The observed spectrum was normalized to the expected rate in the energy interval 0.5--1.5\,k\eve, obtained from the full simulation of the source (known activity within $\pm$5\%) and the setup with MCNP~\cite{mcnp}. As the Compton scattering spectrum from the source is approximately constant in this energy range and the scattered electrons are distributed uniformly in the bulk of the CCD, we take the normalized spectrum as a direct measurement of the device's detection efficiency. The result is in good agreement with the expectation from the simulation of events with a uniform spatial and energy distribution in the bulk of the device (Fig.~\ref{fig:co57}), as used to estimate the detection efficiency to WIMP interactions in the fiducial region.

The rejection factor for surface background in the fiducial region was estimated by simulating events from the front and back surfaces of the CCD.
The gray markers in Fig.~\ref{fig:sige} show the \sxy\ versus energy for one of these simulations, where the interactions were simulated to occur $<$15\,\um\ from the front and back surfaces of the device in the \obo\ data.
The rejection factor is $>$95\% for surface electrons with energy depositions $>$1.5\,k\eve\ and for external photons with incident energies 1.5--4.5\,k\eve. The rejection factor decreases for higher energy photons to 85\% at 6.5\,k\eve\ due to their longer absorption length.
Below 1.5\,k\eve\ the \sxy\ reconstruction worsens, leading to significant leakage into the fiducial region which must be accounted for.

We developed a model of the radioactive background that includes contributions from both bulk and surface events.
We estimated the relative fractions of surface and bulk events in the background from the  \sxy\  distribution of clusters with energies in the range 4.5--7.5\,k\eve, where the expected contribution from a WIMP signal is smallest in the search range.
We used all available data to perform this estimate, including data acquired with a lower gain for  $\alpha$-background spectroscopy studies and excluded from the WIMP search and evaluated background contributions for each CCD individually. 
We estimated that 65$\pm$10\% (60$\pm$10\%) of the total background originated from the CCD bulk (i.e., from Compton scattered photons), 15$\pm$5\% (25$\pm$5\%) from the front, and 20$\pm$5\% (15$\pm$5\%) from the back of the CCD in the \obo~(\obh ) data set.
This background composition was assumed to be energy independent, which is justified by the fact that the background continuum of both bulk and surface events is expected and observed to be approximately constant in energy intervals the size of the WIMP search region. 

\begin{figure}[t!]
\centering
\includegraphics[width=0.48\textwidth]{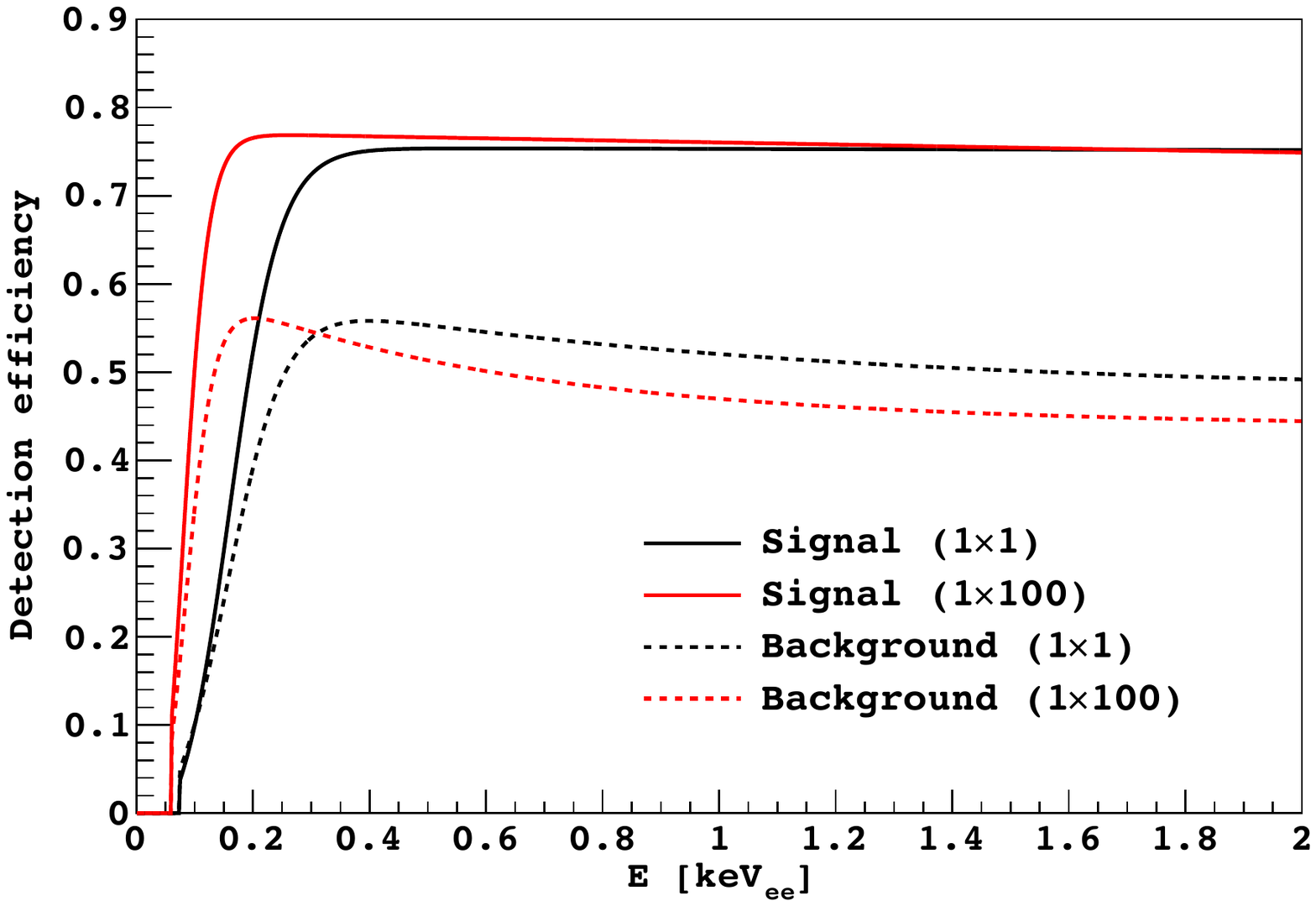}
\caption{Final detection efficiency of events as a function of reconstructed energy ($E$) for the \obo\ (black) and \obh\ (red) data sets after cluster selection outlined in Secs.~\ref{sec:erecsel} and.~\ref{sec:fidu}. Solid lines present the acceptance of the WIMP signal, while dashed lines present the detection efficiency of background events considering both bulk and surface contributions.} \label{fig:eff}
\end{figure}

Fig.~\ref{fig:eff} shows the final detection efficiency after fiducial selection for signal (i.e. WIMP-induced) and background events assuming the initial composition given above.
The turn-on of the efficiency curves near threshold is due to the selection criteria to reject white noise (Sec.~\ref{sec:erecsel}).
At high energies, the signal detection efficiency is almost constant at $\sim$75\%, and the background detection efficiency is dominated by the contribution from Compton events.
The maximum of the background detection efficiency occurs immediately above threshold due to leakage of surface background events.

\section{\label{sec:stat}Likelihood analysis of the spectrum}

After event selection, 31 (23) final candidates remained in the fiducial region with energies $<$7\,k\eve\ in the \obo~(\obh ) data set. Each reconstructed candidate is characterized by its measured electron-equivalent energy, $E_i$. We used this observable to define an extended likelihood function for the signal$+$background model:
\begin{equation*}
\mathcal{L}_{s+b} (s, b, M |\overrightarrow{E})= e^{-(s + b)} \times \prod_{i=1}^{N} \left[ s f_s(E_i|M) + b f_b(E_i) \right],
\label{eq:lsb}
\end{equation*}
where $s$ and $b$ are the expected number of signal and background events in the fiducial region, $f_s(E|M)$ and $f_b(E)$ are the probability density functions (PDFs) for the signal and background, and $N$ is the number of selected events in the data set.

The PDF for the expected WIMP spectrum $f_s(E|M)$ depends on the WIMP mass $M$, the standard halo parameters, and the detector response (ionization efficiency, detection efficiency, and energy resolution):
\begin{eqnarray}
f_s(E|M) &= C(\sigma_{0}) \epsilon_{\rm det}(E) \int \frac{\diff R(E_\textrm{nr}, M, \sigma_{\chi-\textrm{n}}=\sigma_{0})}{\diff E_\textrm{nr}} \nonumber \\*
                 &\times \left|\frac{\diff E_\textrm{nr}}{\diff E_\textrm{ee}}\right| \textrm{Gaus}(E-E_\textrm{ee}, \sigma_\textrm{res}) \diff E_\textrm{ee},
\label{eq:fsPDF}
\end{eqnarray}
where $\diff R(E_\textrm{nr}, M, \sigma_{\chi-\textrm{n}}$$=$$\sigma_{0})/\diff E_\textrm{nr}$ is the predicted WIMP energy spectrum for a reference WIMP-nucleon cross section $\sigma_{0}$, and $C(\sigma_{0})$ is a normalization constant such that the integral of $f_s$ in the search region is normalized to 1.
The nuclear recoil ionization efficiency $E_\textrm{nr}(E_\textrm{ee})$ was used to convert the WIMP energy spectrum, which is a function of the the nuclear recoil energy $E_\textrm{nr}$, to the ionization produced by the nuclear recoil, $E_\textrm{ee}$ (Sec.~\ref{sec:energy}). To account for the finite energy resolution of the detector, we computed the convolution between the $E_\textrm{ee}$ spectrum and a Gaussian distribution with variance $\sigma_\textrm{res}^2$ as modeled in Sec.~\ref{sec:erecsel}. As a last step, the spectrum was multiplied by the detector efficiency for the signal $\epsilon_{\rm det}(E)$ as computed in Sec.~\ref{sec:fidu} (solid lines in Fig.~\ref{fig:eff}).
The PDF for the background $f_b(E)$ is also normalized to 1, and its shape is given by a flat Compton scattering energy spectrum multiplied by the background efficiency (dashed lines in Fig.~\ref{fig:eff}).

To account for performance differences between the \obo\ and \obh\ data sets, we defined a joint likelihood function,
\begin{equation*}
\mathcal{L}_\textrm{joint} (s_\textrm{tot}, \overrightarrow{b}, M |\overrightarrow{E})= \prod_{k=1}^2 \mathcal{L}_{k} (\alpha_k(M) s_\textrm{tot}, b_k, M |\overrightarrow{E}),
\label{eq:jointlike}
\end{equation*}
where the index $k$ runs over the two different data sets, and $\mathcal{L}_{k}$ is the corresponding likelihood function. Note that the functional forms of $f_s$ and $f_b$ depend on the data set as the efficiencies differ between data sets (Fig.~\ref{fig:eff}). The total number of expected signal events $s_\textrm{tot}$ relates to the expected number of events on the $k$th data set through the multiplicative factor $\alpha_k$ that depends on the relative size of the exposure and the signal spectrum from a WIMP of mass $M$.

To quantify the statistical significance of a discovery or to compute an upper limit on the WIMP interaction rate, we performed a hypothesis test based on the profile likelihood ratio statistic $q$. 
This test compares the goodness of fit of two models, one of which, $\mathcal{L}_\textrm{restricted}$, is a special case of the other, $\mathcal{L}_\textrm{free}$.

For this discovery test, the $q$ statistic can be written as
\begin{equation*}
q = -\text{ln} \left[ \dfrac{\text{max}\{ \mathcal{L}_\textrm{restricted}(\overrightarrow{b}|\overrightarrow{E}, s_\textrm{tot} = 0)\}} {\text{max}\{\mathcal{L}_\textrm{free}(s_\textrm{tot}, \overrightarrow{b}, M |\overrightarrow{E})\}}  \right],
\label{eq:qdisc}
\end{equation*}
where the numerator $\text{max}\{\mathcal{L}_\textrm{restricted}\}$ is the maximum value of the likelihood function obtained from a restricted fit with constraints $b_\textrm{\obo~(\obh )}$$>$0 and $s_\textrm{tot}$$=$0, i.e., the null (background-only) hypothesis.
The denominator corresponds to the global maximum obtained from the fit to the data with all parameters free.
The statistic $q$ is positive by construction and values closer to zero indicate that the restricted fit has a likelihood similar to the unconstrained (free) case. On the other hand, large values reflect that the restricted case is unlikely.
To quantify how likely a particular value of $q$ is, the corresponding PDF is required. To compute this distribution, we used a fully frequentist approach and obtained the PDF by performing the estimation of $q$ outlined above for a large number of Monte Carlo samples generated from the background-only model ($s_\textrm{tot}$$=$0).

\begin{figure}[t!]
\centering
\includegraphics[width=0.48\textwidth]{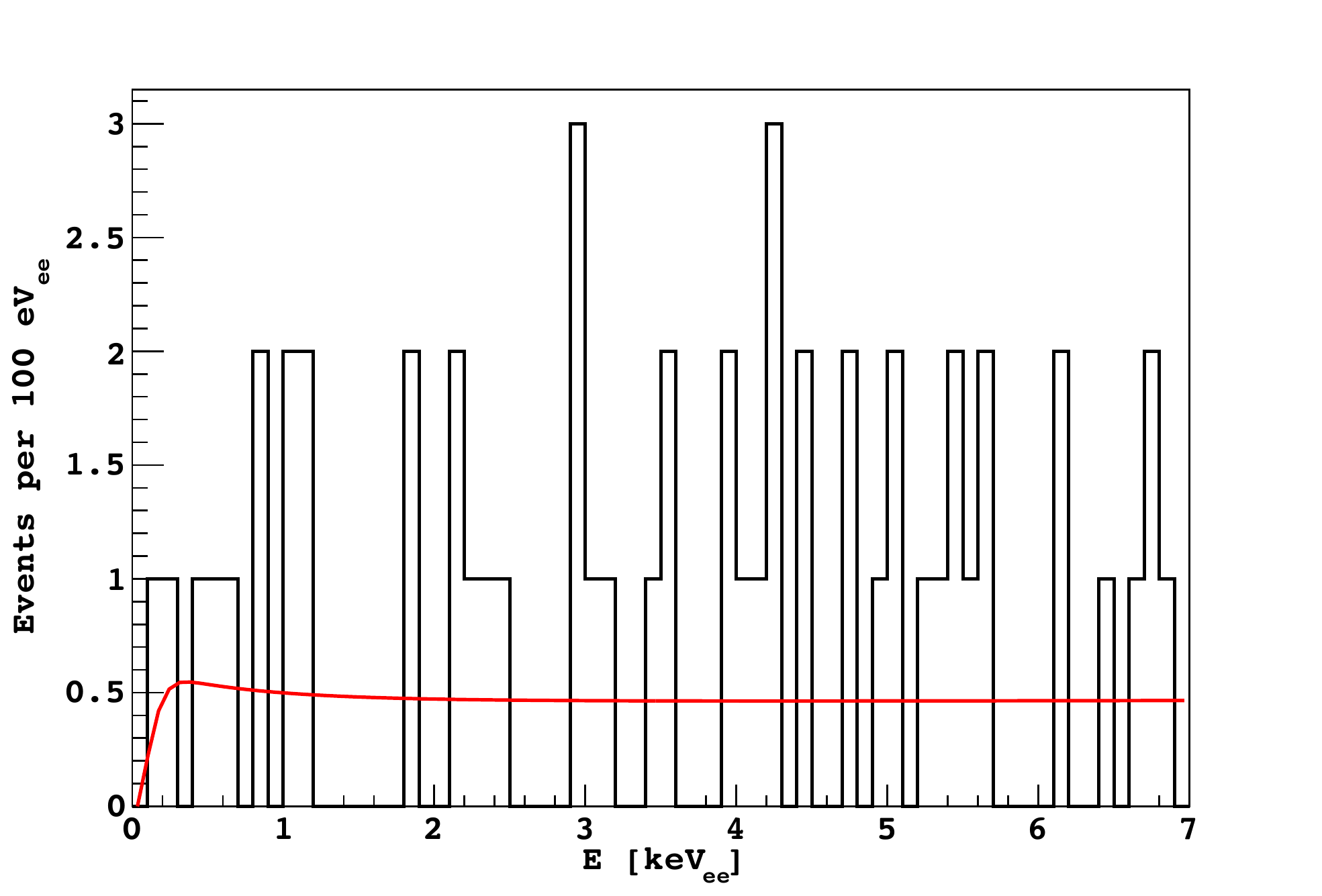}
\caption{Energy spectrum of the final candidates in the \obo\ and \obh\ data sets.
The red line shows the best fit model with parameters $s_\textrm{tot}=0$, $b_\textrm{1x1}=31$ and $b_\textrm{1x100}=23$.} \label{fig:bestfit}
\end{figure}

We performed the discovery test on the joint data set assuming the standard halo parameters:
galactic escape velocity of 544\,km\,s$^{-1}$, most probable galactic WIMP velocity of 220\,km\,s$^{-1}$, mean orbital velocity of Earth with respect to the Galactic center of 232\,km\,s$^{-1}$, and local dark matter density of 0.3\,\gev\,cm$^{-3}$.
We found the recorded events to be compatible with the background-only hypothesis with a p value of 0.8 (Fig.~\ref{fig:bestfit}).
The result corresponds to a dominant background from Compton scattering of 15$\pm$3\,\dru\ (21$\pm$4\,\dru ) in the \obo~(\obh ) data set.

We proceeded to set a 90\% confidence level upper limit on the WIMP-nucleon elastic-scattering cross section, $\tilde{\sigma}_{\chi-\textrm{n}}$.
To compute the upper limit, we followed an analogous approach where, for each value of $M$, we performed a scan on $s$ to find a $\tilde{s}$ such that the test based on the corresponding $q(\tilde{s})$,
\begin{equation*}
q(\tilde{s}) = -\text{log} \left[ \dfrac{\text{max}\{ \mathcal{L}_\textrm{restricted}(\overrightarrow{b}|\overrightarrow{E}, M, s_\textrm{tot} = \tilde{s})\}} {\text{max}\{\mathcal{L}_\textrm{free}(s_\textrm{tot}, \overrightarrow{b}|\overrightarrow{E}, M)\}}  \right],
\label{eq:qlimit}
\end{equation*}
rejected the hypothesis $s_\textrm{tot}$$\ge$$\tilde{s}$ with the desired 90\% C.L.
Note that for each of the scanned masses, we generated the corresponding $q(s)$ distribution from Monte Carlo simulations.

The limit on the WIMP-nucleon cross section $\tilde{\sigma}_{\chi-\textrm{n}}$ was computed from $\tilde{s}$, the total exposure of the experiment $\mathcal{E}$, and the normalization constant $C$ [Eq.~(\ref{eq:fsPDF})] as
\begin{equation*}
\tilde{\sigma}_{\chi-\textrm{n}} = C \frac{\tilde{s}}{\mathcal{E}}.
\label{eq:lim}
\end{equation*}

\begin{figure}[t!]
\centering
\includegraphics[width=0.48\textwidth]{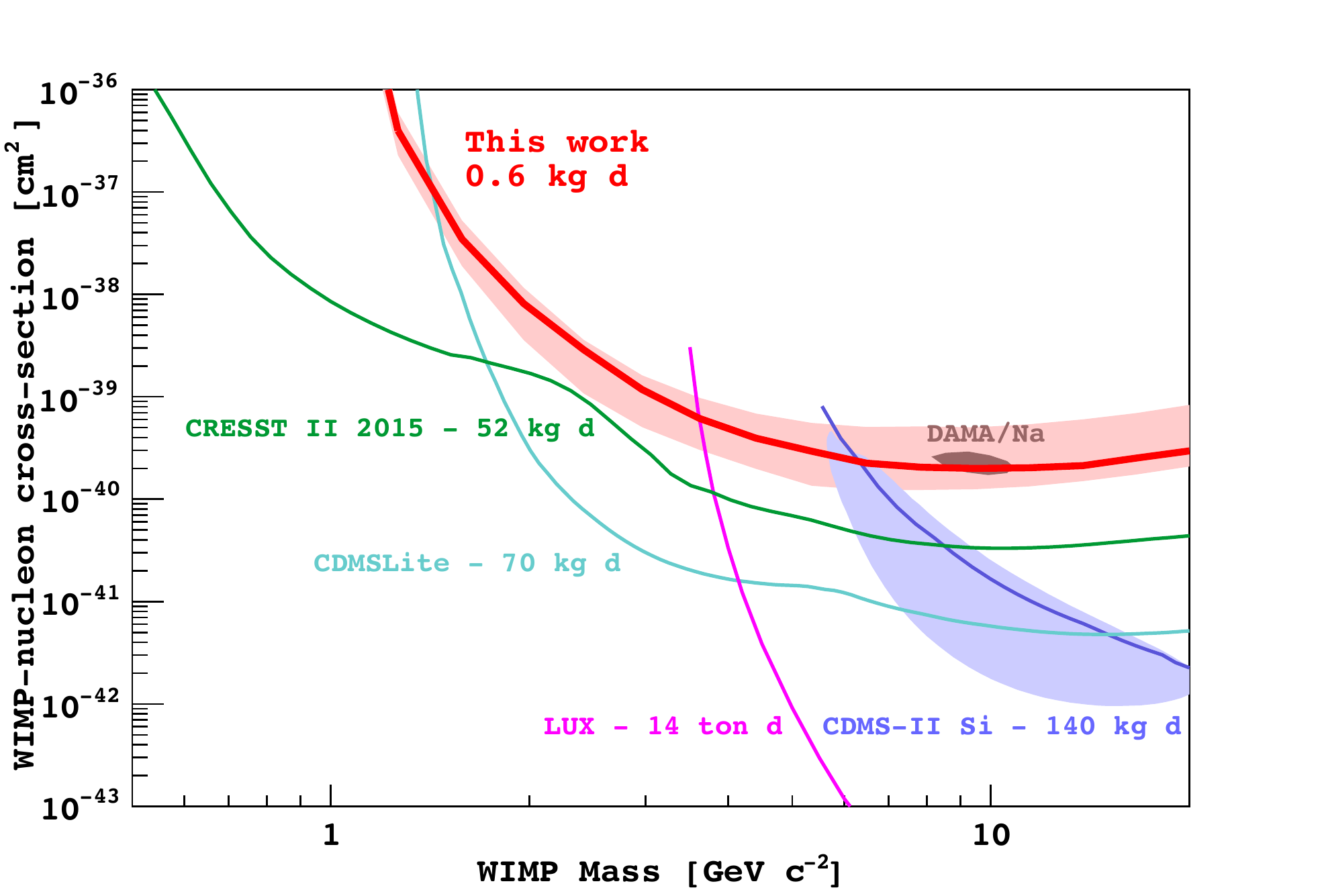}
\caption{Upper limit (90\% C.L.) on the WIMP-nucleon cross section $\tilde{\sigma}_{\chi-\textrm{n}}$ derived from this analysis (red line). The expected sensitivity $\pm$1\,$\sigma$ is shown by the red band. For comparison, we also include 90\% C.L. exclusion limits from other experiments~\cite{Agnese:2013rvf, Angloher:2015ewa, *Agnese:2015nto, *Akerib:2015rjg} and the 90\% C.L. contours corresponding to the potential WIMP signals of the CDMS-II Si~\cite{Agnese:2013rvf} and DAMA~\cite{Savage:2009mk} experiments.}
\label{fig:limThis}
\end{figure}

The 90\% exclusion limit obtained from our data is shown by the red line in Fig.~\ref{fig:limThis}. The wide red band presents the expected sensitivity of our experiment generated from the distribution of outcomes of 90\%~C.L. exclusion limits from a large set of Monte Carlo background-only samples. The good agreement between the expected and achieved sensitivity confirms the consistency between the likelihood construction and experimental data.

The presented limit is particularly robust at low WIMP masses against astrophysical uncertainties in the galactic dark matter velocity distribution because the low threshold of the detector provides sensitivity to interactions from a wide range of WIMP speeds. For example, 15\% and 45\% of all interactions from 3\,\gev\ and 5\,\gev\ WIMPs, respectively, would satisfy the criteria to select ionization events (Sec.~\ref{sec:erecsel}) and produce a signal above electronic noise in DAMIC.

Several sources of systematic uncertainty were investigated. The Fano factor, which is unknown for low-energy nuclear recoils, was varied from 0.13, as for ionizing particles, up to unity.
Exclusion limits were generated changing the nuclear recoil ionization efficiency within its uncertainty~\cite{Chavarria:2016xsi}.
The detection efficiency curves for the signal and the background (Fig.~\ref{fig:eff}) were varied within their respective uncertainties, including those associated to the background composition (Sec.~\ref{sec:fidu}).
All these changes had a negligible impact on the exclusion limit for WIMP masses $>$3\,\gev .
At lower masses, the nuclear recoil ionization efficiency becomes relevant, its uncertainty resulting, for example, in a change by a factor of $\pm$1.5 in the excluded cross section at 2\,\gev .

\section{\label{sec:conc}Conclusion}

We have presented results of a dark matter search performed with a 0.6\,kg\,d exposure of the DAMIC experiment at the SNOLAB underground laboratory. The silicon CCDs employed for the search were extensively characterized, with their ionization response measured down to a threshold of 60\,\eve . The devices operated with remarkably consistent readout noise, allowing for efficient selection of low-energy ionization events. Thanks to the unique spatial resolution of the CCDs, we established the correlation between the spatial extent of a pixel cluster and the depth of the corresponding particle interaction in the silicon substrate. We exploited this information to reject background events occurring near the surfaces of the devices. A total of 54 candidate events were found below 7\,k\eve\ with an energy spectrum consistent with radiogenic backgrounds, and 90\% C.L. exclusion limits on the spin-independent WIMP-nucleon elastic-scattering cross section were derived. To obtain the exclusion limits, we used the most recent measurements of nuclear recoil ionization efficiency in silicon, which cover the relevant energy range down to threshold. A region of parameter space of the potential signal from the CDMS-II Si experiment is excluded using the same nuclear target for the first time.
Even if limited by the exposure and the level of radiogenic background \textemdash\ both to significantly improve in the upcoming DAMIC100 \textemdash\ these results demonstrate DAMIC's sensitivity in the low-mass WIMP region ($<$10\,\gev ), where the experiment is particularly competitive thanks to its low energy threshold and the relatively low mass of the silicon nucleus. In addition, this work firmly establishes the calibration and performance of the detector, the understanding of backgrounds, and the analysis techniques necessary for DAMIC100.

\begin{acknowledgments}
We thank SNOLAB and its staff for support through underground space, logistical and technical services. SNOLAB operations are supported by the Canada Foundation for Innovation and the Province of Ontario Ministry of Research and Innovation, with underground access provided by Vale at the Creighton mine site.
We are very grateful to the following agencies and organizations for financial support: Kavli Institute for Cosmological Physics at the University of Chicago through Grants No. NSF PHY-1125897 and No. PHY-1506208 and an endowment from the Kavli Foundation; Fermi National Accelerator Laboratory (Contract No. DE-AC02-07CH11359); Institut Lagrange de Paris Laboratoire d'Excellence (under Reference No. ANR-10-LABX-63) supported by French state funds managed by the Agence Nationale de la Recherche within the Investissements d'Avenir program under Reference No. ANR-11-IDEX-0004-02; Swiss National Science Foundation through Grant No. 200021\_153654 and via the Swiss Canton of Zurich; Mexico's Consejo Nacional de Ciencia y Tecnolog\'{i}a (Grant No. 240666) and  Direcci\'{o}n General de Asuntos del Personal Acad\'{e}mico - Universidad Nacional Aut\'{o}noma de M\'{e}xico (Programa de Apoyo a Proyectos de Investigaci\'{o}n e Innovaci\'{o}n Tecnol\'{o}gica Grants No. IB100413 and No. IN112213); Brazil's Coordena\c{c}\~{a}o de Aperfei\c{c}oamento de Pessoal de N\'{\i}vel Superior, Conselho Nacional de Desenvolvimento Cient\'{\i}fico e Tecnol\'{o}gico, and  Funda\c{c}\~{a}o de Amparo \`{a} Pesquisa do Estado de Rio de Janeiro.

\end{acknowledgments}

\bibliography{myrefs.bib}

\end{document}